\documentclass[rmp,preprint,showpacs,showkeys,citeautoscript,amsfonts,amssymb,amsmath]{revtex4}

\bibpunct{}{}{,}{s}{}{\textsuperscript{,}}

\newsymbol\vacio 203F
\renewcommand{\Re}{\ensuremath{\mathbb R}}
\newcommand{\N}{\ensuremath{\mathbb N}}

\newcommand{\E}[1]{\ensuremath{\mathbb{E}\!\left[#1\right]}}
\newcommand{\Eb}[1]{\ensuremath{\mathbb{E}\!\left\{#1\right\}}}

\newcommand{\norm}[1]{\ensuremath{\|#1\|}}

\newcommand{\abs}[1]{\ensuremath{\left|#1\right|}}
\newcommand{\sabs}[1]{\ensuremath{|#1|}}

\newcommand{\vect}{{\mathbf r}}

\newcommand{\eZ}{\check{\mathbf{e}}_{z}}
\newcommand{\LR}{\ensuremath{L^2_{\phi}(\Re)}}
\newcommand{\etal}{\textit{et~al.}}

\begin{document}

\title{The  fractional Brownian motion property of the turbulent refractive within Geometric Optics}

\date{\today}

\author{Dar\'{\i}o G. P\'erez}
\renewcommand{\thefootnote}{\emph\alph{footnote}}
\email{dariop@ciop.unlp.edu.ar }
\thanks{This work was carried out while visiting the Mathematisches Institut at the Friedrich-Schiller-Universit\"at in Jena, Germany.}

\affiliation{Centro de Investigaciones \'Opticas (CIOp), CC. 124 Correo Central, 1900 La Plata, Argentina.}

\begin{abstract} 
We introduce fractional Brownian motion processes (fBm) as an alternative model for the turbulent index of refraction. These processes allow to reconstruct most of the refractive index properties, but they are not differentiable. We overcome the apparent impossibility of their use within the Ray Optics approximation introducing a  \textit{Stochastic Calculus}. Afterwards, we successfully provide a solution for the stochastic ray-equation; moreover, its implications in the statistical analysis of experimental data is discussed. In particular, we analyze the dependence of the averaged solution against the characteristic variables of a simple propagation problem.
\end{abstract}
\maketitle

\section{Introduction}

Diverse experimental techniques have been devoted to the study of the optical properties of the turbulent atmosphere. Most of these techniques are based on the analysis of the output of laser beams making their way through it. But also, controlled experiences had been developed for the laboratory, such as the experiments performed by Consortini \etal \cite{consortini1,consortini2,consortini3}. These experiences apply Geometric Optics to interpret the acquired data. This analysis has its theoretical grounds on the precursor paper by P. Beckman \cite{paper:beckman}, who was able to find a simple relationship between the variance of the turbulent refractive index $\mu(\vect)$---being homogeneous and isotropic---and the variance of the  laser beam wandering over a screen. This derivation, however, is based upon the strong assumption that the stochastic process is smooth enough to allow continuity in its derivatives. 

This assumption has serious problems when checked against Tatarsk\u{\i}'s foundational work about lightwave propagation through turbulent atmosphere \cite{book:tatarskii}. A short revision of the covariance of the turbulent index of refraction found by him shows that the stochastic process associated is nowhere differentiable \cite{book:cramer}.

However, the model introduced by Beckmann is just one of the many used during the last 40 years to solve imaging problems through the atmosphere, with relative success. In fact, it is clear that the stochastic properties of the refractive index in Atmospheric Optics have never been fully understood nor explained.  For instance, most works treated it like a Gaussian process \cite{paper:rytov,paper:dashen}, some others suggested stationary increments, while other works have proposed the use of non-Gaussian statistics.  Ishimaru's book \cite{book:ishimaru} presents an extensive description of these works.

Simultaneously, over the last decade an intense debate in Fluid Dynamics has been carried out about whether or not  \textit{passive scalar fields}, among which is the turbulent refractive index, behave like the velocity field. That is, under which circumstances they inherit the stochastic properties of the turbulent velocity $\mathbf{u}(\vect)$, which is a Gaussian process that follows the Kolmogorov refined similarity hypotheses inside the inertial range, $l_0\ll\norm{\vect}\ll L_0,$
\begin{equation}
\E{\norm{\Delta \mathbf{u}(\vect)}^n}= A_n \E{\varepsilon_r^{n/3}} \norm{\vect}^{n/3}, 
\label{eq:KRH}
\end{equation}
where $\varepsilon_r$ is the dissipated energy, $A_n$ is a constant depending on $n$; also, $l_0$ (\textit{inner length}) and $L_0$ (\textit{outer length}) are constants that determine the inertial range. They can be estimated theoretically. When intermittence effects are noticeable the dissipated energy modifies slightly the right-hand side of this equation, changing the power over the correlation distance
\begin{equation*}
\frac{n}{3}\to \frac{n}{3} + \zeta_n \qquad
\end{equation*}
here $\zeta_n$ is called multi-fractal exponent.

It was shown \cite{fred,wang} that passive scalars fields are nearly gaussian as far as $l_0\ll L_0$, and are unsatisfactory modeled either by log-normal distributions, or the so called Frisch's \mbox{$\beta$-model}. Effectively, in the inertial range, these fields obey a law resembling the Kolmogorov's law for the velocity fields. Moreover, they do not present the multi-fractal property due to intermittence whenever an isotropic velocity field is present \cite{mydla}, i.e. $\zeta_n\equiv 0$. Hence, the behavior of the turbulent refractive index predicted by Tatarsk\u{\i} has been confirmed and extended.  

In the meantime, Stolovitzky and Sreenivasan \cite{stolov} successfully obtained Eq.~(\ref{eq:KRH}) modeling the turbulent velocity field as a \textit{fractional Brownian motion} (fBm). But this model failed to replicate the intermittent property. It must be stressed that Kolmogorov refined similarity hypotheses also implies that the velocity field is independent of the dissipated energy probability distribution. 

According to what we have pointed out here, the fBm processes seem to be a good alternative model for the turbulent refractive index. First, they are gaussian, and second, they let us test the Structure Function's power factor. Also, they are continuos but nowhere differentiable, as it results from the application of the Kolmogorov hypotheses to the refractive index. In few words, the fractional Brownian motion model for the turbulent index of refraction describes closer the turbulent properties of passive scalar fields than Beckmann's proposed model. However, a mathematical difficulty is introduced since we have lost differentiability in the usual sense, and as a consequence we cannot appeal to the variational methods used in Geometric Optics. Our task here will be to show how using \textit{Stochastic Calculus} techniques (for a first introduction we refer the reader to {\O}ksendal \cite{oks1,book:holden}) this situation can be overridden providing an explicit solution to the ray propagation through air in turbulent motion. Besides the intrinsic complexity of these tools, our model is meant to provide a bridge between the stochastic processes and experimental data analysis. Also, we will gain knowledge about problems handling (geodesic) stochastic equations in Optics. 

\section{Stochastic Differential Equations in Geometric Optics}


As it is well known\cite{synge,car}, the Lagragian inherited from the application of the Fermat's Extremal Principle in Geometric Optics is singular. In the Appendix \ref{apx:ray-eq} we show the ray-equations associated to this Lagragian result to be, Eqs. (\ref{eq:aceler}):
\begin{equation*}
\frac{d}{d\tau}\left(\frac{1}{\lambda}\frac{d \dot{q}}{d\tau}\right)=\frac{\lambda}{2}\nabla_q n^2(q(\tau)),\quad\text{and the constraint equation } \norm{q}^2 = \lambda^2 n^2(q),
\end{equation*}
where $q(\tau):\Re\rightarrow \Re^3$ is the ray-light trajectory with parameter $\tau$, $n$ is the refractive index, while the smooth function $\lambda: \Re^3\rightarrow \Re$ can be freely choosen. The election of this function fixes the parametrization. 

Since the equations above are nonlinear, we will begin our study by linearizing them. Moreover, at the same time we will set the parameter $\tau$. Let $n$ be the refractive index of the medium and $n_0$ its average, we write
\begin{equation}
n^2(q)= n_0^2 + \alpha \epsilon(q),
\label{index}
\end{equation}
$\epsilon(q)$ represents a perturbation field with its intensity measured by $\alpha$. We also assume that it contains all the inhomogeneities of the media, so when $\alpha=0$ the index is constant. Now we express the solution to Eq.~(\ref{eq:aceler}) in a power series on $\alpha$:
\begin{equation*}
q(\tau)= q_0 + \alpha q_1 + \alpha^2 q_2 + \cdots. 
\end{equation*}

Although we can also develop a series for the constraint function $\lambda$, it is far more convenient to set its value beforehand. Let us rewrite the first equation in Eq.~(\ref{eq:aceler}) as follows
\begin{equation}
\frac{d^2 q}{d\tau^2}= \frac{1}{2}\left[\alpha\lambda^2 \nabla_q\epsilon + \frac{1}{\lambda^2}\left(\nabla_q\lambda^2 \cdot \dot{q}\right) \frac{d q}{d\tau}\right].
\label{eq:start}
\end{equation}
From all the possible parameterization we choose Eq.~(\ref{eq:arclength}), so
\begin{equation*}
\lambda^2(q)= \frac{1}{n^2(q)}= \frac{1}{n_0^2} +\sum^{\infty}_{n=1}\frac{(-1)^n \epsilon^n(q)}{n_0^{2n+2}}\; \alpha^n,
\end{equation*}
in short we will write $\lambda^2_n := (-1)^n \epsilon^n(q)/n_0^{2n+2}$.

Inserting the former series for $\lambda^2$ in (\ref{eq:start}), after some algebraic manipulation we obtain the following family of differential Eqs.~
\begin{align}
\frac{d^2 q_0}{d\tau^2}&=0,\nonumber \\
\frac{d^2 q_1}{d\tau^2}&=\frac{1}{2n_0^2}\left[\nabla_q\epsilon - \left(\nabla_q\epsilon\cdot\frac{d q_0}{d \tau}\right)\frac{d q_0}{d\tau}\right],\quad \text{ and\;when\;}n>1:\label{eq:linsol}\\
\frac{d^2 q_n}{d\tau^2}&=\frac{1}{2}\left\{\lambda^2_n\; \nabla_q\epsilon - \sum^n_{m=1}\left[\sum^m_{k=1} \lambda^2_{k-1} \left(\nabla_q\epsilon\cdot\frac{d q_{m-k}}{d\tau}\right)\right]\frac{d q_{n-m}}{d\tau}\right\},\nonumber
\end{align}
the constraint condition also gives us a constraint equation for each differential equation in Eq.~(\ref{eq:linsol});
\begin{equation*}
\begin{aligned}
\left(\frac{d q_0}{d\tau}\right)^2&=1,\\
\frac{d q_1}{d\tau}\cdot\frac{d q_0}{d\tau}&=0, \\
\sum^{n}_{k=0} \frac{d q_k}{d\tau}\cdot\frac{d q_{n-k}}{d\tau}&=0,\qquad\text{ for\;all\;} n>1.
\end{aligned}
\end{equation*}

We can readily find the zero-order solution of the first equation in (\ref{eq:linsol}). The result is the linear relationship: $q_0(\tau)=\mathbf{a}\,\tau +\mathbf{b}$. Given that the boundary condition to this problem is
\begin{equation}
q(0)=0,
\label{bc}
\end{equation}
it implies that $\mathbf{b}=0$. Now we use the constraint condition to obtain $\norm{\mathbf{a}}^2=1$, so we are free to choose the coordinate frame best suited to our purposes. Let this coordinate frame be:
\begin{equation*}
z\eZ:= q_0 = \tau \eZ,
\end{equation*}
this will be our \textit{forward} direction of propagation. To solve the remaining equations we  need a bit more than algebra.

The turbulent refractive index measures the separation between the index of refraction and its average; $\mu(\vect):=n(\vect)-n_0$. It is a small quantity, that is why its increments  are often replaced in the literature by those of the \textit{permitivity}. This passive scalar field follows the Kolmogorov refined hypotheses in the inertial range, and so its Structure Function is
\begin{equation}
D_{\varepsilon}(r)= \E{\left(n^2(\vect+\vect')-n^2(\vect')\right)^2}= 4 C^2_{\varepsilon}l_0^{2/3-2H}\norm{\vect}^{2H},
\label{eq:index}
\end{equation}
$C^2_{\epsilon}$ is the permitivity structure constant and $H$ is some positive constant less than one. If the turbulence is isotropic and homogeneous then the Kolmogorov hypotheses sets $H=1/3$; so, we have introduced the inner scale to correct the departure from this ideal situation. Then, according to what we have just said at the introduction and the definition (\ref{index}), we propose the following:
\begin{equation}
\epsilon(\vect):= B^H\!\left(l_0^{-1}\norm{\vect}\right)
\label{eq:fBm}
\end{equation}
$B^H$ is a fractional Brownian motion, and $H\in(0,1)$ is the \textit{Hurst parameter}. It is a Gaussian process with the following properties\cite{paper:mandelbrot-2}:
\begin{align*}
\E{B^H(s)}&=0,\\
\E{B^H(s)B^H(t)}&= \frac{1}{2}\left[\abs{s}^{2H}+\abs{t}^{2H}-\abs{s-t}^{2H}\right],
\end{align*}
and the \textit{self-similarity property} 
\begin{equation}
B^H(\alpha s) \overset{d}{=}\alpha^H B^H(s), \quad \text{ for\; any\, }\alpha,
\label{prop:selfsimilar}
\end{equation}
this last equation implies that both variables have the same probability distribution. Thus from Eqs.~(\ref{eq:index}) and (\ref{eq:fBm}) we have,
\begin{align*}
D_{\varepsilon}(r)&= 4 C^2_{\varepsilon}l_0^{2/3-2H}\norm{\vect}^{2H}\nonumber\\
&=\alpha^2\E{\left(B^H(l^{-1}_0\norm{\vect+\vect'})-B^H(l_0^{-1}\norm{\vect'})\right)^2}\nonumber\\
&\simeq\alpha^2l^{-2H}_0\norm{\vect}^{2H},
\end{align*}
whenever $\norm{\vect}\ll\norm{\vect'}$, when, 
\begin{equation*}
\alpha= 2\, l^{1/3}_0\sqrt{ C^2_{\varepsilon}}.
\end{equation*}
Estimates for the \textit{Structure Constant} and the inner length tell us that $\alpha\sim 10^{-6}$. Therefore, in order to examine the stochastic behavior of a wandering beam will be enough to consider the first order solution. 

The first order constraint condition reads then
\begin{equation*}
\frac{d q^z_1}{d z}=0,
\end{equation*}
and together with boundary condition (\ref{bc}) makes the component along the $z$-axis null all over the ray trajectory. Of course, this condition agrees with the corresponding dynamical Eq.~(\ref{eq:linsol}). So we are left with a differential equation for the \textit{perpendicular displacements} to the  direction of propagation. Finally, we multiply these displacements by $\alpha$, and write the first-order equation as
\begin{equation}
\frac{d^2}{d z^2}\,Q= \frac{\alpha}{2 n_0^2} \nabla_q\epsilon\left(z\eZ + Q\right).
\label{eq:non-linear}
\end{equation}

We must provide a context to understand the previous equation. That is, a stochastic equation is not only determined by the type of process (the fractional Brownian motion in our case) attached to it, but also by the \textit{integro-differential} theory employed to define its \textit{derivatives}. Moreover, there are distinctive stochastic integration methods whether $H>1/2$ or $H\leq 1/2$ \cite{decreusefond}. Here we are going to make use of the Stochastic Calculus exposed in the Appendix \ref{apx:fsc}, so only the $H>1/2$ case will be considered. By doing so, either we are considering the inertial-diffusive range, in the following sense
\begin{equation*}
\zeta_n>\frac{1}{3},
\end{equation*}
or the anisotropic scalar situation $\zeta_n \rightarrow 1$ \cite{paper:elperin}. This situation could be observed in a laboratory if an isotropic velocity field can not achieved by the experimental setup. 

Because the turbulent refractive index oscillates around its mean value, it is expected that the light wanders around the \mbox{$z$-axis} over the screen. So the solution we are looking for must have expectation zero. This can easily achieved by the formalism we are employing: the stochastic integrals (formally known as \textit{ fractional It\^o integrals}) defined by the fractional white noise and Wick product on fractional Hida spaces have expectation zero. Henceforth, from definition (\ref{eq:fBm}) we can calculate the gradient of the index of refraction. We have, using the continuity and differential properties for the fractional Brownian motion---Eqs.~(\ref{fBm-cont}) and (\ref{fWn-cont})---in $\mathcal{S}^*_H$ and applying the chain rule, the following
\begin{equation*}
\frac{\partial}{\partial x^i}\left[B^H(l_0^{-1}\norm{\vect})\right]=\left.\frac{dB^H}{d s}\right|_{s=l_0^{-1}\norm{\vect}}\!\!\!\times \frac{x^i}{l_0\norm{\vect}}= \frac{W^H\left(l_0^{-1}\norm{\vect}\right)}{l_0\norm{\vect}}\,x^i, \, i=1,2,
\end{equation*}
where $W^H$ is the fractional white noise. Remember, once more, this last identity must be understood in terms of the formal definition of white noise inside the fractional Hida spaces, it has nothing to do with the usual concept of derivative. 

Next, the procedure to interpret Eq.~(\ref{eq:non-linear}) requires to replace all the ordinary products containing stochastic variables by Wick products, and we finally write
\begin{equation}
\frac{d^2}{d z^2}\,Q= \frac{\alpha}{2\, l_0n_0^2}\ \left[\frac{W^H\!\!\left(l_0^{-1}\norm{z\eZ + Q}\right)}{\norm{z\eZ + Q}}\right] \diamond Q.
\label{volterra}
\end{equation}
The fractional white noise is a functional on the real line and its composition with another stochastic process has to be defined. Because any analytic function is expressed by a power series, as {\O}ksendal \etal\cite{book:holden} suggest, we follow our substitution rule for products and replace the powers in the series by Wick's powers---just as we did with the Wick exponential in Eq.~(\ref{eq:wick}) from the Appendix \ref{apx:fsc}---whenever a stochastic process is an argument for the given function. The representation for the noise in $\mathcal{S}^*_H$ is a series with analytic functions as components, Eq.~(\ref{eq:white-noise}); thus, we change these components 
\begin{equation*}
\int_{\Re} \phi(s,z)\,\tilde{\xi}_k(s)\, ds\rightarrow \int_{\Re} \phi^{\diamond}(s,Z)\,\tilde{\xi}_k(s)\,ds,
\end{equation*}
$Z$ is some continuous stochastic process with $\E{Z}:=z_0\neq 0$, and $\phi^{\diamond}(s,\cdot)$ is the Wick representation of $\phi(s,\cdot)= H(2H-1) \abs{s- \cdot}^{2H-2}$. We can write $\norm{z\eZ+Q}\simeq z + Q^2/2z$ because $Q\sim \mathcal{O}(\alpha)$, and then we can evaluate the fractional white noise at $z +\alpha^2 Z(\omega)$:
\begin{equation*}
\phi^{\diamond}(s,z +\alpha^2 Z)= H(2H-1)\abs{z +\alpha^2 Z-s}^{\diamond(2H-2)}=H(2H-1)\left[(z-s)+\alpha^2 Z\right]^{\diamond(2H-2)},
\end{equation*}
we have just took the positive part of the absolute value; it is enough for us examine this situation. If $\E{\alpha^2 Z}=\alpha^2 z_0$ then
\begin{multline*}
\left[(z-s)+\alpha^2 Z\right]^{\diamond(2H-2)}\\
= \left[(z-s)+ \alpha^2 z_0\right]^{2H-2} +
 \sum^{\infty}_{n=1}  \frac{\alpha^{2n}(2H-2)\cdots(2H-3-n)}{n!\left[(z-s)+ \alpha^2 z_0\right]^{n+2-2H}}\left(Z-z_0\right)^{\diamond n},
\end{multline*}
and all the terms in the series are of order higher or equal to 2 in $\alpha$. We just need to compare the first term against the deterministic coefficient in the white noise series,
\begin{equation*}
\phi(s,t+\alpha^2 z_0)-\phi(s,t)\sim (t-s)^{(2H-3)}\alpha^2
\end{equation*}
and because this happens coordinate to coordinate in the fractional white noise decomposition we find
\begin{equation*}
\frac{W^H\!(l_0^{-1}z)}{z} - \frac{W^H\!\!\left(l_0^{-1}\norm{z\eZ + Q}\right)}{\norm{z\eZ + Q}}
\sim \mathcal{O}(\alpha^2).
\end{equation*}
The first-order Eq.~(\ref{volterra}) is unaffected by this replacement since they differ in $\alpha^2$. We have arrived at the linear equation,
\begin{equation}
\frac{d^2}{d z^2}\,Q(z)= g \; \frac{W^H\!\!\left(l_0^{-1}z\right)\diamond Q(z)}{z},
\label{eq:diff-vol}
\end{equation}
we have set $g=\alpha/2\, l_0 n_0^2$. 

\section{The Stochastic Volterra Equation}

\subsection{The Stochastic Volterra Equation and Its Solution}

The integral form of Eq.~(\ref{eq:diff-vol}) is,
\begin{equation}
\begin{aligned}
Q(z)&= \dot{Q}_0 z + g \int^z_0\int^s_0 \frac{W^H(l_0^{-1}s')}{s'}\diamond Q(s')\: ds' ds\\
&=\dot{Q}_0 z + g \int^z_0 \frac{(z-s)}{s} W^H(l_0^{-1}s)\diamond Q(s)\:ds.
\end{aligned}
\label{eq:volterra}
\end{equation}
Let us set the following initial conditions $Q(0)=0$ and $\dot{Q}(0)\in \mathcal{S}_H^{\ast}$. We are interested in finding a solution on the interval $0\leq z\leq L$. What we have here is a stochastic Volterra equation with (Fredholm) kernel
\begin{equation*}
k^H(z,s):= g \frac{(z-s)}{s} \chi_{[0,z]}(s) W^H(l_0^{-1} s).
\end{equation*}
which is continous and
\begin{equation}
\norm{k^H(z,s)}_{H,-q}\leq g \tilde{M}\; \chi_{[0,z]}(s)s^{-1}\abs{z-s},
\label{eq:notbounded}
\end{equation}

Now we have to find the conditions that make Eq.~(\ref{eq:volterra}) solvable. We propose as \textit{ansatz} the usual resolvent for convoluted kernels, that is,
\begin{equation*}
K_H(z,s)=\sum^{\infty}_{n=1} K^{(n)}_H(z,s),
\end{equation*}
such that 
\begin{align}
Q(z)&= \dot{Q}_0 z + \int^z_0 K_H(z,s)\diamond \left(\dot{Q}_0 s\right) ds\nonumber\\
&=\dot{Q}_0 \diamond\left[ z + \int^z_0 K_H(z,s) s \,ds\right]
\label{eq:solution}
\end{align}
with the $K^{(n)}_H$ given inductively by
\begin{align}
K^{(n+1)}_H(z,s)&= \int^z_s K^{(n)}_H(z,u)\diamond k^H(u,s)\; du,\quad \text{ with\quad }n\geq 1,\\
K^{(1)}_H(z,s)&= k^H(z,s).
\label{eq:kernel-al}
\end{align}
{\O}ksendal\cite{book:holden} found this is the unique solution for bounded kernels in the distribution Hida space. The same procedure can also be applied to fractional Hida spaces. But it can be seen from Eq.~(\ref{eq:notbounded}) that it is not our case when $s\rightarrow 0$. Nevertheless, a new norm can be defined so that under it the former resolvent exists. Let it be,
\begin{equation*}
\norm{k^H}_{L^{p,p'}(J),-q}= \sup_{\substack{\norm{F}_{L^p(J),-q}\leq 1\\\norm{G}_{L^{p'}(J),-q}\leq 1}} \int_J\int_J \norm{G(z)k^H(z,s)F(s)}_{H,-q}\, dz\, ds
\end{equation*}
where $\norm{F}_{L^p(J),-q}=\left(\int_J \norm{F(s)}^p_{H,-q}\,ds\right)^{1/p}$ is the Lebesque integral, and $J=(0,L]$. This norm is simplified using the H\"older inequality the following relation can be proved:
\begin{equation}
\begin{aligned}
\norm{k^H}_{L^{p,p'}(J),-q}&\leq \min\left\{\left[\int_J\left(\int_J\norm{k^H(z,s)}^p_{H,-q}ds\right)^{p'/p} dz\right]^{1/p'}\!\!\!,\right.\\
&\qquad \left.\left[\int_J\left(\int_J\norm{k^H(z,s)}^{p'}_{H,-q}dz\right)^{p/p'} ds\right]^{1/p}\right\}.
\end{aligned}
\label{cond:norm}
\end{equation}
Gripenberg \etal \cite{grip} discuss the deterministic counterpart of this construction. They proved that a resolvent solution exists whenever the norm of the kernel is less than one. This theorem can be tracked back to our norm in the stochastic case. Hence, the same hypothesis applies for this stochastic Fredholm kernel: $\norm{k^H}_{L^{p,p'}(J),-q}< 1$ for some $q>0$. Then applying Eq.~(\ref{eq:notbounded}) to the bounding condition (\ref{cond:norm}) we find 
\begin{equation*}
\norm{k^H}_{L^{p,p'}(J),-q}\leq g \tilde{M} p^{-1/p} \left(\frac{\pi p'}{\sin \pi p'}\right)^{1/p'}<1,
\end{equation*}
since $\tilde{M}$ is a small constant and $g\ll 1$. This guarantees the convergence of the proposed ansatz. 

The solution represented as a series of convoluted kernels, Eqs.~(\ref{eq:solution})--(\ref{eq:kernel-al}), is useless for calculations. Next, we will prove that a fractional chaos expansion exists for the solution. Let us take the second term in the Wick product of Eq.~(\ref{eq:solution}), it can be written 
\begin{equation*}
X(z)= z+ \int^z_0 \left[\sum^{\infty}_{n=1} K_H^{(n)}(z,s)\right] s\; ds =z+ \sum^{\infty}_{n=1}\left[\int^z_0 K_H^{(n)}(z,s) s\; ds\right],
\end{equation*}
because it converges absolutely. The general term in this series can be written,
\begin{align*}
\int^z_0 K_H^{(n)}(z,s) s\; ds& \\
&= g l_0^{1-H}\!\!\int^z_0\!\left[\int^z_{s_1} K_H^{(n-1)}(z,s_2)\frac{(s_2-s_1)(s_1-0)}{s_1}\,ds_2\right]\diamond  W^H(s_1)\,ds_1\nonumber\\
&= g l_0^{1-H}\!\!\int^z_0\!\int^z_{s_1} K_H^{(n-1)}(z,s_2)\frac{(s_2-s_1)(s_1-0)}{s_1}\,ds_2\,dB^H_{s_1}\nonumber\\
&= (g l_0^{1-H})^2\!\!\int^z_0\!\!\int^z_{s_1}\!\!\int^z_{s_2}\!\! K_H^{(n-2)}(z,s_3)\,\frac{(s_3-s_2)(s_2-s_1)(s_1-0)}{s_2s_1}\,ds_3\,dB^H_{s_2}dB^H_{s_1}\nonumber\\
&= (g l_0^{1-H})^n\!\!\int^z_0\cdots\int^z_{s_{n-1}}\int^z_{s_n}\frac{(z-s_n)(s_n-s_{n-1})\cdots(s_1-0)}{s_n s_{n-1}\cdots s_1}\,dB^H_{s_n}\cdots dB^H_{s_1}\nonumber\\
&= (g l_0^{1-H})^n\!\!\int_{\Re_+^n}(z-s_n)\prod^{n}_{i=1}\left[\frac{(s_i-s_{i-1})}{s_i} \chi_{[s_{i-1},z)}(s_i)\right]\,dB^H_{s_n}\cdots dB^H_{s_1}\nonumber\\
&= z(g z^H l_0^{1-H})^n\!\!\int_{\Re_+^n}f^{(n)}(s_n,\dots,s_1)\,dB^H_{s_n}\cdots dB^H_{s_1}.
\end{align*}
We have used property (\ref{prop:selfsimilar}) to build the above adimensional integrals, and defined 
\begin{equation}
f^{(n)}(s_n,\dots,s_1)=(1-s_n)\prod^{n}_{i=1}\left[\frac{(s_i-s_{i-1})}{s_i} \chi_{[s_{i-1},1)}(s_i)\right].
\label{eq:chaos-term}
\end{equation}
with $s_0=0$. Now, the symmetrized form $\hat{f}^{(n)}(s_n,\dots,s_1)=(1/n!)\sum_{\sigma\in\Pi}f^{(n)}(s_{\sigma_n},\dots,s_{\sigma_1})$ induces the following relation
\begin{equation*}
\int_{\Re_+^n}\!\hat{f}^{(n)}(s_n,\dots,s_1)\,dB^H_{s_n}\cdots dB^H_{s_1}=\!\int_{\Re_+^n}\!\!f^{(n)}(s_n,\dots,s_1)\,dB^H_{s_n}\cdots dB^H_{s_1},
\end{equation*}
and finally,
\begin{equation}
X(z)=z\left\{ 1+ \sum^{\infty}_{n=1}\int_{\Re^n_+}\left[\tilde{g}^n\, \hat{f}^{(n)}(s_n,\dots,s_1)\right]dB^H_{s_n}\cdots dB^H_{s_1}\right\},
\label{eq:chaos-exp}
\end{equation}
where $\tilde{g}= l_0g (z/l_0)^H$. This will be nothing else but the fractional chaos expansion provided
\begin{equation*}
\sum^{\infty}_{n=1}\tilde{g}^{2n} \sabs{\hat{f}^{(n)}}^2_{\phi}<\infty
\end{equation*}
holds. In fact this condition express nothing else that the existence of the variance of the process,
\begin{equation*}
\mathbb{E}(X^2_z)= z^2 \left[1+ \sum^{\infty}_{n=1} \tilde{g}^{2n} \sabs{\hat{f}^{(n)}}^2_{\phi}\right],
\end{equation*}
we used property (\ref{eq:orthonorm-int}). The search for an upper bound for the succession of $\phi$-norms, given that the $\hat{f}^{(n)}$ are symmetric, is straightforward:
\begin{multline}
\sabs{\hat{f}^{(n)}}^2_{\phi}= \int_{\Re^{2n}_+}\! \hat{f}^{(n)}(s_n,\dots, s_1)\hat{f}^{(n)}(s'_n,\dots,s'_1)\phi(s_n,s'_n)\cdots\phi(s_1,s'_1)\, ds_n\cdots ds_1\,ds'_n\cdots ds'_1\leq\\
\leq \int^1_0\int^1_0\!\!\cdots\!\!\int^1_{s_{n-1}}\int^1_{s'_{n-1}}\phi(s_n,s'_n)\cdots\phi(s_1,s'_1)\, ds_n\cdots ds_1\,ds'_n\cdots ds'_1,
\label{eq:bound2}
\end{multline}
because of definition (\ref{eq:chaos-term}) and the fact $0<s_i-s_{i-1}\leq s_i$ (idem $0<s'_i-s'_{i-1}\leq s'_i$) the last inequality follows. Observing that
\begin{align*}
\int^1_{s_{n-1}}\int^1_{s'_{n-1}}\phi(s_n,s'_n) ds_n\,ds'_n&\\
&= H(2H-1)\int^1_{s_{n-1}}\int^1_{s'_{n-1}}\abs{s_n-s'_n}^{2H-2}ds_n\,ds'_n\\
&= \frac{1}{2}\left[(1-s_{n-1})^{2H}+(1-s'_{n-1})^{2H}-\abs{s_n-s'_n}^{2H}\right]\leq 1,
\end{align*}
we iteratively apply it in Eq.~(\ref{eq:bound2}) to find: $\sabs{\hat{f}^{(n)}}^2_{\phi} \leq 1.$ Thus, the chaos expansion exists for all $ z\leq L$ whenever
\begin{equation*}
l_0 g \left( \frac{L}{l_0}\right)^H<1
\end{equation*}
is satisfied.  From the definition of $g$ and the magnitude of the quantities 
involved in it we have: 
\begin{equation}
L\ll l_0^{1-1/3H} (C^2_\varepsilon)^{-1/2H}.
\label{eq:order-of}
\end{equation}
So, the condition above is always fulfilled.

\subsection{Ray-light Statistics: An Example}

In this section we will use the stochastic ray-equation solution to study the statistical properties of the turbulent refractive index. Both coordinates of displacement are independent, and they also hold the same (non-coupled) differential equation. There is enough to consider the 1-dimensional then. The parameter election, Eq.~(\ref{eq:arclength}), we have used in our treatment, also defines the meaning of the \textit{transversal} velocities, for they are the angles of deviation. Being the velocities continuous we can set,
\begin{equation*}
\dot{Q}_0:=\lim_{\epsilon \rightarrow 0}\dot{Q}(\epsilon) = \left.\theta\right|_{\epsilon=0}\in \mathcal{S}^*_H.
\end{equation*}
Since our solution is dependent of the initial refractive angle $\theta$, its behavior at the boundary, $\epsilon\rightarrow 0$, must be known.  This boundary is just the interface between turbulent and resting air. Henceforth, we will also model the initial angle as a fractional Brownian motion,
\begin{equation*}
\theta(\epsilon)= c \int_{\Re_+}\chi_{[0,\epsilon)}(s)\, dB^H_s=c\, B^H(\epsilon),
\end{equation*}
the constant $c$ is adimensional and measures the strength of the noise. The length $\epsilon$ works as a kind of correlation distance, as it goes to zero we are examining the properties of the interface's short-range correlation. 

Besides, any stochastic process can be put in terms of the spans described in the Appendix \ref{apx:fsc}, and these depend on the construction of stochastic integrals by step functions. So, even if the former model needs to be corrected---maybe the interface introduces long-range correlations---the next results are useful; since, they are the building blocks for more complex stochastic processes.

The solution (\ref{eq:solution}) is written using the chaos expansion (\ref{eq:chaos-exp}) and the initial conditions:
\begin{equation*}
Q(z)= \theta(\epsilon) \diamond X_z=z c \,B^H(\epsilon)\diamond \left( 1+ \sum^{\infty}_{n=1}\, \tilde{g}^n \int_{\Re^n_+}\!\! \hat{f}^{(n)}(s_n,\dots,s_1)\,dB^H_{s_n}\cdots dB^H_{s_1}\right)\!.
\end{equation*}
From the Wick product properties is easy to see that
\begin{equation*}
\E{Q(z)}=z c\,\E{B^H(\epsilon)}\E{1+ \sum^{\infty}_{n=1}\,\tilde{g}^n \int_{\Re^n_+}\!\! \hat{f}^{(n)}(s_n,\dots,s_1)\,dB^H_{s_n}\cdots dB^H_{s_1}}= 0 \cdot \E{1} = 0.
\end{equation*}

The evaluation of the variance from experimental data is the most common topic in many works related to the optical properties of turbulence because it is directly related to the Structure Constant. Hence, we calculate it using property (\ref{eq:cov}),
\begin{equation*}
\E{Q^2(z)} = c^2  \mathbb{E}(B^H_{\epsilon}\diamond X_z)^2= c^2 \left[\mathbb{E}(D_{\Phi_{\chi_{[0,\epsilon)}}}X_z)^2 + \mathbb{E}(X^2_z) \sabs{\chi_{[0,\epsilon)}}^2_{\phi}\right].
\end{equation*}
$\mathbb{E}(X^2_z)$ was already evaluated in the last section. The fractional Malliavin derivative appearing at the right-hand side demands elaboration, property (\ref{eq:dbrow1}) implies
\begin{equation*}
D_{\Phi_{\chi_{[0,\epsilon)}}} X_z = \int_{\Re_+} (D^{\phi}_s X_z)\, \chi_{[0,\epsilon)}(s)\,ds.
\end{equation*}
Since the $\phi$-differential is linear so we have 
\begin{equation}
D^{\phi}_s X_z= z \sum^{\infty}_{n=1}  \tilde{g}^n  D^{\phi}_s\!\left[ \int_{\Re^n_+}\hat{f}^{(n)}(s_n,\dots,s_1)\, dB^H_{s_n}\cdots dB^H_{s_1}\right].
\label{eq:d-series}
\end{equation}
We are going to compute these derivatives now: let us fix $n\geq 2$, from the first theorem (\ref{thm:first}) we can commute the stochastic integral and $\phi$-differential,
\begin{multline*}
 D^{\phi}_s\!\!\left[ \int_{\Re^n_+}\hat{f}^{(n)}(s_n,\dots,s_1) dB^H_{s_n}\cdots dB^H_{s_1}\right]\\
 = \int_{\Re_+}\!\! D^{\phi}_s\!\left[\int_{\Re^{n-1}_+}\!\hat{f}^{(n)}\; dB^H_{s_n}\cdots dB^H_{s_2}\right]dB^H_{s_1}+ \int_{\Re^n_+}\!\hat{f}^{(n)}\; dB^H_{s_n}\cdots dB^H_{s_2}\phi(s,s_1)ds_1.
\end{multline*}
Now, we recursively commute the operators, the $\phi$-differential and the Wick integral. Each time we do so another integral as the last one on the right-hand side of the equation above is added. After $(n-1)$ iterations we reach the innermost integral, thus we evaluate
\begin{equation*}
D^\phi_s\!\!\left[\int_{\Re_+}\hat{f}^{(n)}(s_n,\dots,s_1)\,dB^H_{s_n}\right]=\int_{\Re_+}\hat{f}^{(n)}(s_n,\dots,s_1)\phi(s_n,s)\,ds_n,
\end{equation*}
with the aid of property (\ref{eq:dbrow2}). Finally, 
\begin{multline*}
D^{\phi}_s\!\!\left[ \int_{\Re^n_+}\hat{f}^{(n)}(s_n,\dots,s_1) dB^H_{s_n}\cdots dB^H_{s_1}\right] =\int_{\Re^{n}_+}\!\hat{f}^{(n)}\; \phi(s,s_n)ds_n dB^H_{s_{n-1}}\cdots dB^H_{s_1}+\\
+\dots+\int_{\Re^{n}_+}\!\hat{f}^{(n)}\; dB^H_{s_{n}}\cdots\phi(s,s_k)\;ds_k \cdots dB^H_{s_1}+\dots+\int_{\Re^n_+}\!\hat{f}^{(n)}\; dB^H_{s_n}\cdots dB^H_{s_2}\phi(s,s_1)ds_1\\
= n \int_{\Re^{n-1}_+}\!\left[\int_{\Re_+}\hat{f}^{(n)}\; \phi(s,s_n)ds_n \right]dB^H_{s_{n-1}}\cdots dB^H_{s_1},
\end{multline*}
to arrive to the last equality the symmetry of $\hat{f}^{(n)}$ was employed. Instead, for $n=1$ we just use property (\ref{eq:dbrow2}):
\begin{equation*}
D^{\phi}_s\!\!\left[ \int_{\Re_+}\!\!\hat{f}^{(1)}(s_1)\,  dB^H_{s_1}\right]=\int_{\Re_+}\!\!f^{(1)}(s_1)\,  \phi(s,s_1)\,ds_1.
\end{equation*}

Afterwards, we can build the fractional Malliavin derivative (\ref{eq:malliavin-derivative}) from the series (\ref{eq:d-series}),
\begin{multline}
 D_{\Phi_{\chi_{[0,\epsilon)}}} X_z\! =  z\tilde{g}\left\{\int_{\Re^2_+}\!\!f^{(1)}(s')\,\chi_{[0,\epsilon)}(s) \phi(s,s')\,ds'\, ds\right.+\\
+\left.\sum^{\infty}_{n=1}\,(n+1)\,\tilde{g}^n\!\!\!
\int_{\Re^n_+}\!\!\left[\int_{\Re^2_+}\!\hat{f}^{(n+1)}(s',\dots,s_1)\;\chi_{[0,\epsilon)}(s) \phi(s,s')\,ds' ds \right]\!dB^H_{s_n}\cdots dB^H_{s_1}\right\}; 
\label{eq:diff-var}
\end{multline}
its second moment is
\begin{multline*}
\E{(D_{\Phi_{\chi_{[0,\epsilon)}}} X_z )^2}= z^2\tilde{g}^2\left\{\left[\int_{\Re^2_+}\!\!f^{(1)}(s')\,\chi_{[0,\epsilon)}(s) \phi(s,s')\,ds'\, ds\right]^2+\right.\\
+\left.\sum^{\infty}_{n=1}\,(n+1)^2\tilde{g}^{2n}\abs{\int_{\Re^2_+}\!\hat{f}^{(n+1)}(s',\cdot)\;\chi_{[0,\epsilon)}(s) \phi(s,s')\,ds' ds\:}^2_{\phi}\right\}.
\end{multline*}
This series converges, we apply the same procedure as before to find a bound for the integrals. What is more, each norm appearing in the series is bounded by the zero term,
\begin{multline*}
\abs{\int_{\Re^2_+}\!\hat{f}^{(n+1)}(s',\cdot)\;\chi_{[0,\epsilon)}(s) \phi(s,s')\,ds' ds\:}^2_{\phi}\leq \left[\int^{\epsilon}_0\int^1_0\!\!(1-s')\phi(s,s')\,ds'\, ds\right]^2 =\\
=\left\{\frac{1}{2(2H+1)}\left[1-\epsilon^{2H+1}-(1-\epsilon)^{2H+1}\right]-\frac{H}{2H+1}\,\epsilon^{2H+1}+\frac{1}{2}\right\}^2\leq \frac{1}{4}.
\end{multline*}
Thus, the existence of Eq.~(\ref{eq:diff-var}) is guaranteed. 
Finally, the variance of the displacements, with the norm
\begin{equation*}
\sabs{\chi_{[0,\epsilon)}}_{\phi}^2= H(2H-1)\int^{\epsilon}_0\!\!\int^{\epsilon}_0 \abs{u-s}^{2H-2}du\,ds= {\epsilon}^{2H},
\end{equation*}
is written as,
\begin{multline*}
\E{Q^2(z)}=z^2 c^2 \left\{ \left[ \epsilon^{2H} + \tilde{g}^2\left(\int_{\Re^2_+}\!f^{(1)}(s')\;\chi_{[0,\epsilon)}(s) \phi(s,s')\,ds' ds\:\right)^{\!\!2}\right] + \right.\\
+\left.\sum^{\infty}_{n=1} \,\tilde{g}^{2n}\!\! \left[\epsilon^{2H}\sabs{\hat{f}^{(n)}}^2_{\phi}+(n+1)^2\tilde{g}^2\abs{\int_{\Re^2_+}\!\hat{f}^{(n+1)}(s',\cdot)\;\chi_{[0,\epsilon)}(s) \phi(s,s')\,ds' ds\:}^2_{\phi}\right]\right\}.
\end{multline*}

Now, as the correlation distance goes to zero we recover the initial condition. While terms coming from the second moment of $X_z$ banish (they are all bounded and multiplied by $\epsilon^{2H}$), it is not the case with those coming from the fractional derivative. We will not go through copious calculations since we are interested in a general outline of the solution; thereof, the solution can be expressed as
\begin{multline*}
\E{Q^2(z)}\\
=\frac{z^2 c^2 \tilde{g}^2 }{4}
+ z^2 c^2 \tilde{g}^2 \sum^{\infty}_{n=1} \,\tilde{g}^{2n}(n+1)^2 \left.\abs{\int_{\Re^2_+}\!\hat{f}^{(n+1)}(s',\cdot)\;\chi_{[0,\epsilon)}(s) \phi(s,s')\,ds' ds\:}^2_{\phi}\right|_{\epsilon\rightarrow 0}.
\end{multline*}
We can estimate a bound for the second term:
\begin{align*}
F(\tilde{g}^2)=4\sum^{\infty}_{n=1} \,\tilde{g}^{2n}(n+1)^2 \Bigg|\int_{\Re^2_+}\!\hat{f}^{(n+1)}(s',\cdot)\;\chi_{[0,\epsilon)}(s) \phi(s,s')&\left.\,ds' ds\:\Bigg|^2_{\phi}\right|_{\epsilon\rightarrow 0} \\
\leq & 4\sum^{\infty}_{n=1} \,\tilde{g}^{2n}(n+1)^2 \left(\frac{1}{4}\right)\\
\leq &\frac{1}{\tilde{g}^2}\sum^{\infty}_{n=2} n^2 \tilde{g}^{2n}.
\end{align*}
Now, because $x/(1-x)^2=\sum_{n=1}^\infty n x^{2n}$ it is
\begin{equation*}
\frac{1}{\tilde{g}^2} \sum^\infty_{n=2} n^2 \tilde{g}^{2n} = \frac{1}{\tilde{g}}\sum_{n=1}^\infty n^2 \tilde{g}^{2n-1}- 1 =\frac{1}{2\tilde{g}}\frac{d}{d\tilde{g}}\!\!\left[\frac{\tilde{g}^2}{(1-\tilde{g}^2)^2}\right] -1= \frac{(1+\tilde{g}^2)}{(1-\tilde{g}^2)^3}-1
\end{equation*}
and then $F(\tilde{g}^2)\leq \tilde{g}^2/(1-\tilde{g}^2)^3$, whenever $\tilde{g}<1$. Finally, replacing the values for $\tilde{g}$, we have
\begin{equation}
\E{Q^2(z)}=\frac{c^2}{4}\,  A\, z^{2H+2} l_0^{2/3-2H} \left[1 + F\left( A\, z^{2H} l_0^{2/3-2H}\right)\right].\label{example-solution}
\end{equation}
Furthermore, for the range of validity given in the past sections, the contribution of the function $F$ is less than $10^{-6}$. Thus, the first contribution to Malliavin derivative of $X$ completely characterize the variance once the interface's properties are defined. So, determining the behavior of the interface is crucial for the present model.
\section{Remarks \& Conclusions}

In this paper we provided a fractional Brownian motion model for the turbulent index of refraction; afterwards, we used this model to build a stochastic ray-equation---a stochastic Volterra equation. For which we have given a (unique) solution. Our analysis just covers the $H>1/2$ case which is meant for non-isotropic or near diffusive range turbulence; that is, the temperature gradients are relevant. 

Besides the fact that these Hurst exponents are hardly found in field experiences. They can easily appear within the laboratory. In particular, the turbulence is assumed \textit{per se} nearly isotropic and homogeneous among optics works even in situations where the turbulence is created but not controlled. The example shown here can be used to test these cases. Also, we pretend to give a glimpse of the $H\leq 1/2$ problem extrapolating these results.

We observe that the solution (\ref{example-solution}) strongly depends on the initial conditions; our election of the initial angle at the example is the natural choice given the behavior of scalar quantities in turbulence. Under this condition, we apply Eq.~(\ref{example-solution}) to estimate the centroid's variance of a laser beam at a distance $L$---for any legth $L$ less than $10^6$m according to Eq.~(\ref{eq:order-of}):
\begin{equation}
\text{Var}[Q(L)]\simeq \frac{c^2}{4} A\, l^{2/3-2H}_0 L^{2+2H},
\label{eq:variance}
\end{equation}
where the correction to this result is of order $\mathcal{O}(F)\sim 10^{-5}$. We must note that this is the contribution from the Malliavin derivative. That is, the constant term, of order zero, does not contribute to the variance. Also, the anisotropy introduced by the termal flux should be observed in different values of $c$ for each axis.

Now, as $H^+\rightarrow 1/2$ the variance of the displacement approaches to $C^2_{\varepsilon} l^{-1/3}_0 L^{3}$. We notice that this behavior is the same found in Consortini \etal\cite{consortini3}. But it not comes from the isotropic Kolmogorov's law, with the hypotheses made at the introduction it corresponds to a Brownian motion ($H=1/2$). That is, given a gaussian process with Structure Function's power as in Eq.~(\ref{eq:KRH}) the result from Consortini does not follow. We have shown this can only be achieved on inertial-diffusive conditions, for $\zeta_n \searrow 1/3$; shortly, the Beckman's hypotheses \cite{paper:beckman} do not apply to gaussian models used in Fluids Dynamics.

Nevertheless, the approach we have introduced is just the beginning of a long journey, since we must examine the stochastic ray-equation when $H\in[1/3,1/2)$. By doing so, we will confirm that the power law for the variance of the displacements is univocally determined and no discontinuities arise, Eq.~(\ref{eq:variance}) is still valid.  But, this will require the introduction of other tools---even new for the Stochastic Analysis---since in this range any \textit{smoothness} property of the fBm processes is lost \cite{paper:errami}, and this inquiry will be the topic of future works.

A step further should be considered afterwards. The proposed model for the turbulent index of refraction, Eq.~(\ref{eq:fBm}), just approximates the Structure Function. After we give a solution to the stochastic ray-equation for the whole range of Hurst parameters with this model, we will start examining other functionals of the fractional Brownian motion, which refines our theoretical results against the observed properties. 

\begin{acknowledgments}

I would like to express my gratitude to the Antorchas Foundation, who support my stay in Jena. To Prof. Dr. Martina Z\"ahle  for the helpful discussions about fractional Stochastic Calculus in the development of this paper and her guidance in this new topic. 

I also extend my thanks to Aicke Hinricks and Ra\'ul Enrique for their useful comments.
\end{acknowledgments}
\appendix
\section{The Ray-Path Equations}\label{apx:ray-eq}

Fermat's Extremal Principle in Geometric Optics, means to find the variational solution to
\begin{equation*}
\delta\!\left( \int n\, ds \right) = 0.
\end{equation*}
The solution to this equation is interpreted as the ray trajectory, which we shall denote here as $q(\tau)$. The parameter $\tau$ has, in principle, no physical meaning. 
In Optics Treatises this parameter is usually replaced with one of the trajectory coordinates, 
which fulfills $dq_i/d\tau>0,$ and is thus called the propagation direction. But the election of this parameter can not be done at will\cite{synge}, since, for any parameterization chosen, the Optical Lagrangian
\begin{equation*}
L(q,\dot{q})=n(q)\norm{\dot{q}},
\end{equation*} 
($q,\dot{q} \in \Re^3$ are the position and velocity respectively) is \textit{degenerated}. Its  solution is not univocally determined because \cite{krupkova}
\begin{equation*}
\text{ det}\left(\frac{\partial^2 L}{\partial \dot{q}^i\partial\dot{q}^j}\right)\equiv 0,
\end{equation*}
for any pair $(q,\dot{q})$. That is, calculating the momentum,
\begin{equation}
p_i=\frac{\partial L}{\partial \dot{q}^i }= n(q)\frac{\dot{q}^i}{\norm{q}},
\label{eq:momentum}
\end{equation}
we can write the Lagrangian as follows,
\begin{equation}
L(q,\dot{q})=\sum_i \frac{\partial L}{\partial \dot{q}^i}\, \dot{q}^i=\sum_i p_i \dot{q}^i.
\label{eq:lag}
\end{equation}
This is homogeneous in the velocities which allows us to recalculate the momentum and find,
\begin{align*}
\frac{\partial L}{\partial \dot{q}^j} =\sum_i \left(\delta^{i j} \frac{\partial L}{\partial \dot{q}^i} + \frac{\partial^2 L}{\partial \dot{q}^i\partial\dot{q}^j} \dot{q}^j\right),\quad\text{ then}\nonumber\\
 0=\sum_i \frac{\partial^2 L}{\partial \dot{q}^i\partial\dot{q}^j}\, \dot{q}^j, \quad \forall\dot{q}^i,
\end{align*}
so this matrix is singular as we stated above. Another consequence can be derived from Eq.~(\ref{eq:momentum}); also, it induces the following relation 
\begin{equation*}
\norm{p}^2 = n^2(q),
\label{eq:condition}
\end{equation*}
which indicates that the choice of coordinates and momenta is not free. 

Now, the degeneracy of the Lagrangian can be worked out in the Hamiltonian framework using the constraint we have just found. This problem of constrained Hamiltonians is known as Dirac's problem in the literature \cite{arnold}. The procedure is as follows: first, define the constraint function \begin{equation*}
\Psi(p,q)=\frac{1}{2}\left[\norm{p}^2-n^2(q)\right]; 
\end{equation*}
then, calculate the Hamiltonian $H$ from the original Lagrangian, from Eqs.~(\ref{eq:momentum}) and (\ref{eq:lag}) it is exactly zero
\begin{equation*}
H = \sum_i p_i \dot{q}^i-L = \sum_i p_i \dot{q}^i- \sum_i p_i \dot{q}^i\equiv 0;
\end{equation*}
finally, build a new Hamiltonian
\begin{equation*}
\widetilde{H}(p,q):= H(p,q)+\lambda \Psi(p,q) = \lambda \Psi(p,q)
\end{equation*}
and apply the variational procedure to all the coordinates included $\lambda$. 

By doing so, we obtain the following dynamic equations
\begin{equation}
\left\{
\begin{aligned}
\dot{p}=& - \frac{\partial \widetilde{H}}{\partial q} = \lambda \frac{\partial \Psi}{\partial q}= \frac{\lambda}{2} \nabla_q n^2\\
\dot{q}=& \frac{\partial \widetilde{H}}{\partial p} = \lambda \frac{\partial \Psi}{\partial p}= \lambda p
\end{aligned}\right.
\label{eq:hamilton}
\end{equation}
and the constraint, over the phase space,
\begin{equation}
0=\Psi(p,q)=\frac{1}{2}\left[\norm{p}^2-n^2(q)\right].
\label{eq:const}
\end{equation}
To ensure $\lambda$ is well defined we need to introduce \textit{compatibility conditions}. Which arise from establishing the constraints as motion constants, that is, $\dot{f}=0=\{f,\widetilde{H}\}$ with $\{\cdot,\cdot\}$ the Poisson bracket. In our problem these conditions reduce just to one: $\{H,\Psi\}=0$ which is automatically accomplished. There are no \textit{secondary constraints} derived from the compatibility conditions so Eqs.~(\ref{eq:hamilton}) and (\ref{eq:const}) completely define our problem \cite{blago}. Notice that $\lambda$ is actually a smooth function on the constrained space that can be freely chosen.

Finally, this pair of equations can be combined to yield
\begin{equation}
\begin{aligned}
\frac{d}{d\tau}\left(\frac{1}{\lambda}\frac{d q}{d\tau}\right)&=\frac{\lambda}{2}\nabla_q n^2(q(\tau))\quad \text{ and}\\
\norm{\dot{q}}^2&= \lambda^2 n^2(q).
\end{aligned}
\label{eq:aceler}
\end{equation}
We now realize that with each selection we make for $\lambda$ the parameter $\tau$ is also set, i.e. if we choose $\lambda=n^{-1}$ then
\begin{equation}
\norm{\dot{q}}^2= 1\qquad\text{ and}\qquad ds=\norm{\dot{q}}d\tau=d\tau 
\label{eq:arclength}
\end{equation}
$\tau$ is then the arc-length. But selecting $\lambda=1$ gives us $ds=n d\tau$ and now the parameter is $\tau=\int ds/n.$

\section{Fractional Stochastic Calculus}\label{apx:fsc}

Here we will introduce briefly the elements needed to build a stochastic calculus for $B^H$. The reader can find a complete reference in Hu and {\O}ksendal \cite{paper:Hu} and Duncan \etal \cite{paper:duncan}. 

Let  $H\in(1/2,1)$ be a fixed constant, and let us define
\begin{equation*}
\phi(s,z)= H(2H-1)\abs{s-z}^{2H-2},\qquad s,z\in\Re.
\end{equation*}
Then $f\in\LR$, if it is measurable and 
\begin{equation}
\abs{f}^2_{\phi}:=\int_{\Re}\int_{\Re} f(s)f(z)\phi(s,z)\,\,ds\, dz <\infty.
\label{cond:squared}
\end{equation}
Also the inner product can be defined in $\LR$;
\begin{equation*}
(f,g)_{\phi}:=\int_{\Re}\int_{\Re} f(s)g(z)\phi(s,z)\,\,ds\, dz,\quad \text{ for\,all\,}f,g\in\LR,
\end{equation*}
and it becomes a separable Hilbert space.

Let $\mathcal{S}(\Re) \subset \LR$  be the Schwartz space of rapidly decreasing smooth functions on $\Re$. From its dual the probability space $\Omega=\mathcal{S}'(\Re)$ can build, it is the space of tempered distributions $\omega$ on $\Re$. Its associated probability measure, $\mu_{\phi}$, can be found applying the \mbox{Bochner-Minlos} theorem,
\begin{equation*}
\E{e^{i\langle\cdot,f\rangle}}:=\int_{\Omega} e^{i\langle\omega,f\rangle} d\mu_{\phi}(\omega)=e^{-\frac{1}{2}\abs{f}_{\phi}^2},
\end{equation*}
where $\langle\omega,f\rangle$ is the usual pairing between elements in the dual and functions on $\Re$. With this definition it is easy to prove that
\begin{equation}
\E{\langle\cdot,f\rangle}=0,\text{ and }\,\E{\langle\cdot,f\rangle^2}=\abs{f}^2_{\phi}.
\label{eq:mean}
\end{equation}
The triplet $\left(\Omega, \mathcal{B}\!\left(\Omega\right),\mu_{\phi}\right)$ is thus a probability space---$\mathcal{B}\!\left(\Omega\right)$ is the Borel algebra on $\Omega$---usually called \textit{fractional white noise probability space}. Then let $L^2(\mu_{\phi})=L^2(\Omega, \mathcal{B}\!\left(\Omega\right),\mu_{\phi})$ be the space of all the random variables $X:\Omega \rightarrow \Re$ such that
\begin{equation*}
\norm{X}_{L^2(\mu_{\phi})}^2:=\mathbb{E}\abs{X}^2<\infty.
\end{equation*}
Hence, those functions in \LR\, define the set of random variables of the form  $f(\omega)=\langle\omega,f\rangle$ which is included in  $L^2(\mu_{\phi})$; that is, the condition (\ref{cond:squared}) induces square measurable random variables because of Eqs.~(\ref{eq:mean}).

It can also be shown that $\mathcal{S}(\Re)$ is dense in $\LR$. This means that for any $f\in\LR$ we can build a series $f_n\in\mathcal{\Re}$ such that $f_n\rightarrow f$ in $\LR$.  What is more, the following limit exists in $L^2(\mu_{\phi})$:
\begin{equation}
\lim_{n\rightarrow\infty}\langle\omega,f_n\rangle:=\langle\omega,f\rangle.
\label{eq:lim}
\end{equation}

We can now define  the fractional Brownian motion process as follows:
\begin{equation}
B^H(z):=B^H(z,\omega)=\langle\omega,\chi_{[0,z)}\rangle \in L^2(\mu_{\phi}).
\label{eq:fbm}
\end{equation}
In this definition $B^H$ is thought to be the $z$-continous version of the rightmost hand side term. The \textit{step function} $\chi_{[0,z)}:\Re\rightarrow [-1,1]$ is defined by
\begin{equation*}
\chi_{[0,t]}(s)=\left\{
\begin{aligned}
1&\quad\text{if }0\leq s\leq t\\
-1&\quad\text{if } t< s\leq 0.\\
0&\quad\text{otherwise}
\end{aligned}\right.
\end{equation*}
So taking $f\in \LR$, approximating it by step functions, and then using property (\ref{eq:lim}) traduces definition (\ref{eq:fbm}) to
\begin{equation}
\langle\omega,f\rangle=\int_{\Re} f_z\, dB^H\!(z,\omega)
\label{eq:intfbm}.
\end{equation}
It can be verified using the same procedure for $f,g\in\LR$ that
\begin{equation}
\E{\langle \omega,f\rangle\langle \omega, g\rangle}= (f,g)_{\phi}
\label{isometry}.
\end{equation}

Again let $f$ be as above, as it is shown in Duncan {\etal} once defined
\begin{equation}
\mathcal{E}(f)= \exp\!\left(\int_{\Re} f\, dB^H-\frac{1}{2}\abs{f}^2_{\phi}\right),
\label{eq:basis}
\end{equation}
the \textit{linear span}
\begin{equation*}
\mathcal{E}=\left\{\sum^n_{k=1} a_k \mathcal{E}(f_k); n\in\N, a_k\in\Re, f_k\in\LR \text{\; for\, }k\in\{1,\dots,n\}\right\},
\end{equation*}
is dense in $L^2(\mu_{\phi})$. 

Also there is another functional expansion from where this Lebesgue space with measure $\mu_{\phi}$ can be build. This expansion is useful, in particular, to introduce some tools we will use through this work. To do so let us define the Hermite functions\cite{book:sundaram} 
\begin{equation*}
\xi_n(x)= \left(2^{n-1} (n-1)! \sqrt{\pi}\right)^{- 1/2} e^{-x^2/2} H_{n-1}(x); \quad n=1,2,\cdots
\end{equation*}
on \Re. This set of functions is an orthonormal basis in $L^2(\Re)$. We can map this basis to an orthonormal one $\tilde{\xi}_n=\Gamma^{-1}_{\phi}\xi_n\in\LR$, by means of the isometry 
\begin{equation*}
\Gamma_{\phi}f(s)=c_H\int_{[s,\infty)}(z-s)^{H-3/2} f(z)\, dz,
\end{equation*}
where
\begin{equation*}
c_H=\sqrt{\frac{H(2H-1)\,\Gamma\!\left(\frac{3}{2}-H\right)}{\Gamma\!\left(H-\frac{1}{2}\right)\Gamma(2-2H)}}.
\end{equation*}
It can also be shown that
\begin{equation*}
\int_{\Re}\tilde{\xi}_n(s)\, \phi(s,z)\,ds =c_H\int_{(-\infty,z]}(z-s)^{H-3/2} \xi_n(s)\,ds,
\end{equation*}
because the $\tilde{\xi}_n$'s are an orthonormal basis these integrals are also smooth.

Let $\mathcal{I}$ be the set of all finite multi-indices $\alpha=(\alpha_1,\cdots,\alpha_m)$ of nonnegative integers, we define
\begin{equation*}
\mathcal{H}_{\alpha}(\omega):= H_{\alpha_1}\!(\langle\omega,\tilde{\xi}_1\rangle)\cdots H_{\alpha_m}\!(\langle\omega,\tilde{\xi}_m\rangle).
\end{equation*} 
In particular, if we let $\varepsilon^i:=(0,\cdots,0,1,0,\cdots,0)$ denote the $i$'th unit vector then we get from Eq.~(\ref{eq:intfbm}) and the definition of Hermite polynomials
\begin{equation*}
\mathcal{H}_{\varepsilon^i}(\omega)= H_1\!(\langle\omega,\tilde{\xi}_i\rangle)=\langle\omega,\tilde{\xi}_i\rangle=\int_{\Re}\tilde{\xi}_i(s)\; dB^H_s.
\end{equation*}
These functionals are elements of $L^2(\mu_{\phi})$, and they form its basis. That is, for $X\in L^2(\mu_{\phi})$ there are $c_{\alpha}\in\Re$ and $\alpha\in\mathcal{I}$, such that
\begin{equation}
X(\omega)= \sum_{\alpha\in\mathcal{I}} c_{\alpha}\mathcal{H}_{\alpha}(\omega),
\label{eq:1series}
\end{equation}
and also
\begin{equation*}
\norm{X}^2_{L^2(\mu_{\phi})}=\sum_{\alpha\in\mathcal{I}}\alpha! c_{\alpha}^2.
\end{equation*}
These coefficients are given by $c_{\alpha}=\E{X\,\mathcal{H}_{\alpha}}/\alpha!$. With this property at hand we now define the fractional Hida spaces: the \textit{fractional Hida test function space} $\mathcal{S}_H$ as the set of all
\begin{equation*}
\begin{aligned}
\psi(\omega)&=\sum_{\alpha\in\mathcal{I}}a_{\alpha}\mathcal{H}_{\alpha}(\omega)\in L^2(\mu_{\phi}),\quad\text{ such\; that}\\
\norm{\psi}^2_{H,k}&=\sum_{\alpha\in\mathcal{I}}\alpha!\, a^2_{\alpha}(2\,\N)^{k\alpha}<\infty,\quad\text{ for\; all }\>k\in\N,
\end{aligned}
\end{equation*}
where 
\begin{equation*}
(2\,\N)^{\gamma}= \prod_j (2j)^{\gamma_j}\quad\text{ for\; any\; element }\>\gamma=(\gamma_1,\cdots,\gamma_m)\in\mathcal{I}.
\end{equation*}
The \textit{fractional Hida distribution space} $S^{\ast}_H$ is the set of all formal expansions
\begin{equation}
\begin{aligned}
Y(\omega)&=\sum_{\beta\in\mathcal{I}}b_{\beta}\mathcal{H}_{\beta}(\omega),\quad\text{ such\; that}\\
\norm{Y}^2_{H,-q}&=\sum_{\beta\in\mathcal{I}}\beta!\, a^2_{\beta}(2\,\N)^{-q\beta}<\infty,\quad\text{ for\; some }\>q\in\N.
\end{aligned}
\label{eq:hida-exp}
\end{equation}
With these definitions is not hard to see that $\mathcal{S}_H\subset L^2(\mu_{\phi})\subset\mathcal{S}_H^{\ast}$.

It is now time to show how the fractional white noise and integration with respect to $B^H$ can be defined. Let us first calculate the expansion for the stochastic integral in Eq.~(\ref{eq:intfbm}). For any $f\in L^2_{\phi}(\Re)$---any given deterministic function---we have from Eqs.~(\ref{eq:1series}) and (\ref{isometry}):
\begin{equation*}
\int_{\Re} f_s\> dB^H_s= \sum^{\infty}_{i=1} (f,\tilde{\xi}_i)_{\phi}\,\mathcal{H}_{\varepsilon^i}(\omega).
\end{equation*}
When $f=\chi_{[0,z)}$ in the left hand side we recover Eq.~(\ref{eq:fbm}) and the following relation holds
\begin{equation*}
B^H(z)= \sum^{\infty}_{k=1}\left[\int_{[0,z)}\left(\int_{\Re}\tilde{\xi}_k(s)\phi(s,u)\,ds \right)du\right]\mathcal{H}_{\varepsilon^k}(\omega)\;\in\mathcal{S}^{\ast}_H,
\end{equation*}
We can check its norm
\begin{equation}
\begin{aligned}
\norm{B^H\!(z)}^2_{H,-q}&=\sum^{\infty}_{k=1}\left[\int_{[0,z)}\left(\int_{\Re}\tilde{\xi}_k(s)\phi(s,u)\,ds \right)du\right]^2(2k)^{-q}\leq \\
&\leq M^2\,z^2\,2^{-q}\sum^{\infty}_{k=1}k^{1/3-q}=M^2\,z^2\,2^{-q}\zeta\!\left(q-\frac{1}{3}\right),
\end{aligned}\label{fBm-cont}
\end{equation}
($\zeta$ is the Riemann's zeta function) this because of
\begin{equation*}
\abs{\int_{\Re}\tilde{\xi}_k(s)\phi(s,u)\, ds}=\abs{\int_{(-\infty,u]}(u-s)^{H-3/2}\,\xi_k(s)\,ds}\leq M\,k^{1/6}
\end{equation*}
according to Szeg\"o \cite{book:szego}. When $q>4/3$ the former inequality shows that $B^H$ is continuos and differentiable in $\mathcal{S}^{\ast}_H$. Hence, 
\begin{equation}
\frac{d}{dz}B^H(z)= \sum^{\infty}_{k=1} \left(\int_{\Re}\tilde{\xi}_k(s)\phi(s,z)\,ds \right)\mathcal{H}_{\varepsilon^k}(\omega):=W^H(z)\in \mathcal{S}^{\ast}_H \label{eq:white-noise}
\end{equation}
is the formal definition of fractional white noise. Again it is also continous, when \mbox{$z>s$}
\begin{align}
\norm{W^H(z)-W^H(s)}^2_{H,-q}&=\sum^{\infty}_{k=1}\varepsilon^i!\left|\int_{\Re}\tilde{\xi}_k(u)\phi(z,u)du-\int_{\Re}\tilde{\xi}_k(u)\phi(s,u)du \right|^2(2k)^{-q}\leq\nonumber\\
&\leq c_H \sum^{\infty}_{k=1} \left[\int^{z-s}_0\!\! [(z-s)-u]^{H-3/2}\abs{\xi_k(s+u)}du\right]^2(2k)^{-q}\leq\nonumber\\
&\leq 2^{-q} c_H^2  M^2\,\zeta\!\left(q+\frac{1}{6}\right) \left\{\int^{z-s}_0 [(z-s)-u]^{H-3/2}\right\}^2=\nonumber\\
&=\frac{2^{2-q} c_H^2  M^2}{(2H-1)^2}\,\zeta\!\left(q+\frac{1}{6}\right) (z-s)^{2H-1}.
\label{fWn-cont}
\end{align}

We now define the Wick product, like in Hu and {\O}ksendal, as follows: let be $X(\omega)=\sum_{\alpha\in\mathcal{I}} a_{\alpha} \mathcal{H}_{\alpha}(\omega)$ and $Y(\omega)=\sum_{\beta\in\mathcal{I}} b_{\beta} \mathcal{H}_{\beta}(\omega)$ in $\mathcal{S}^{\ast}_H$ then
\begin{equation}
(X\diamond Y)(\omega)=\sum_{\alpha,\beta\in\mathcal{I}} a_{\alpha}b_{\beta}\mathcal{H}_{\alpha+\beta}(\omega)=\sum_{\gamma\in\mathcal{I}}\left(\sum_{\alpha+\beta=\gamma}a_{\alpha}b_{\beta}\right)\mathcal{H}_{\gamma}(\omega),
\label{eq:wick-prod}
\end{equation}
this product is commutative, associative and distributive like the usual product in \Re. Under the norm $\norm{\cdot}_{H,-q}$ it can be shown that all the power series defined in the real line have their counterpart in the distribution Hida space. If we understand by $X^{\diamond n}$ the $n$-times Wick product of $X\in\mathcal{S}^{\ast}_H$, then the \textit{Wick exponential} is    
\begin{equation*}
\exp^{\diamond}(X)=\sum^{\infty}_{n=0}\frac{1}{n!}X^{\diamond n}.
\end{equation*}
In particular when $X=\langle \omega,f\rangle=\int_{\Re}f_s dB^H_s$ holds, then
\begin{equation}
\exp^{\diamond}(\langle \omega,f\rangle)=\mathcal{E}(f)
\label{eq:wick}.
\end{equation}
The right-hand side was defined in Eq.~(\ref{eq:basis}), and therefore we have shown the link between the two representations presented up to now. 

It is appropriate to introduce the fractional Malliavin derivative or $\phi$-derivative, for $X\in L^2(\mu_{\phi})$ and $g\in L^2_{\phi}(\Re)$, 
\begin{equation*}
D_{\Phi g}X(\omega)=\lim_{\delta\rightarrow 0} \frac{1}{\delta}\left\{X(\omega + \delta \int_{\Re}
(\Phi g)(u)\,du))-X(\omega)\right\}
\end{equation*}
where $(\Phi g)(z)=\int_{\Re}\phi(z,u)g_u\,du$. Afterwards, if there exists a function $D^{\phi}_s X$ such that
\begin{equation}
D_{\Phi g}X =\int_{\Re} (D^{\phi}_s X)\,g_s\,ds,\> \forall g\in \LR,
\label{eq:malliavin-derivative}
\end{equation}
we say that $X$ is $\phi$-differentiable, and $D^{\phi}_s X$ is the $\phi$-differential. These are differential operators, and they also present the following properties: let $X$ be as always and $f,g:\Re\rightarrow\Re$ then
\begin{align}
D_{\Phi g}f(X)=f'(X)D_{\Phi g}X,\nonumber\\
D_{\Phi g}\langle\omega,f\rangle=(f,g)_{\phi},\label{eq:dbrow1}\\
D^{\phi}_s \langle\omega,f\rangle= \int_{\Re} \phi(u,s) f_u du\label{eq:dbrow2}
\end{align}

Moreover, it can be proved for any two $X, Y\in L^2(\mu_{\phi})$---decomposed by the span $\mathcal{E}$---using the Wick product and the $\phi$-derivative properties that,
\begin{equation}
\Eb{\left[X\diamond\!\int_{\Re}f_s\,dB^H_s\right]\!\left[Y\diamond\!\int_{\Re}g_s\,dB^H_s\right]}=\E{(D_{\Phi f}X)( D_{\Phi g}Y) + XY(f,g)_{\phi}}.\label{eq:cov}
\end{equation}

These properties allows to change the integrator inside Eq.~(\ref{eq:intfbm}) by a stochastic function $X: \Re \times \Omega \rightarrow \Re$. The basic procedure consists of building a Riemann sum, replacing the standart product by the Wick one, 
\begin{equation}
S_n(X) := \sum_{i=0}^{n-1} X_{z_i}\diamond (B^H_{z_{i+1}}-B^H_{z_i}),
\label{eq:riemannsum}
\end{equation}
and finally observing its convergence (whenever $\mathbb{E}\abs{X}^2_{\phi}<\infty$) under the $\mathbb{E}\abs{\cdot}^2_{\phi}$-norm. Moreover, the following equality holds
\begin{equation*}
\int_{[0,L)}\!X_s\> dB^H_s =\int_{[0,L)}\! X_s \diamond W^H_s\> ds,
\end{equation*}
while the integral on the left-hand side represents the limit of Eq.~(\ref{eq:riemannsum}), the right-hand side is just the integral evaluated under the Hida expansion of the Wick product defined in Eqs.~(\ref{eq:hida-exp})--(\ref{eq:wick-prod}).

We will complete this Appendix enumerating some useful theorems from Duncan \etal \cite{paper:duncan}:

\textit{Let $X_z$ a stochastic process defined as above, and $\sup_{z\in [0,L)} \mathbb{E}\abs{D^{\phi}_z X}^2_{\phi}<\infty$. Also, let $\eta_z=\int_{[0,z)}f_s\, dB^H_s$. Then for $s,z\in [0,L)$}
\begin{equation*}
D^{\phi}_s \eta_z =\int_{[0,z)}\!D^{\phi}_s X_u\, d B^H_u + \int_{[0,z)}\! X_u \phi(s,u)\, du.
\label{thm:first}
\end{equation*}

\textit{Let $\eta_z=\int_{[0,z)}f_s\, dB^H_s$ and $F(z,x):\Re_+\times\Re\rightarrow \Re$, where $f$ is continous and $F$ has second continous derivatives. Then}
\begin{equation}
\begin{aligned}
F(z,\eta_z)&= F(0,0)+\int_{[0,z)} \frac{\partial F}{\partial s}(s,\eta_s)\, ds + \int_{[0,z)}\frac{\partial F}{\partial x} (s,\eta_z) f_s\, dB^H_s\\
&+ \int_{[0,z)}\frac{\partial^2 F}{\partial x^2}(s,\eta_s) \int_{[0,s)} \phi(s,s') f_{s'}\, ds'ds.
\end{aligned}\label{eq:itof}
\end{equation}

\textit{If $X\in L^2(\mu_{\phi})$ then there exists a sequence $\{f_n \in L^2_{\phi}(\Re^n_+)\}_{n\in \N}$ such that $\sum^{\infty}_{n=1} \abs{f_n}^2_{\phi}< \infty$ and}
\begin{equation*}
X= \E{X} + \sum^{\infty}_{n=1} \int_{\Re_+} f_n(s_1,\cdots,s_n)\> dB^H_{s_1}\cdots dB^H_{s_n}.
\end{equation*}
Here it is understood $L^2_{\phi}(\Re^n_+)$ as the $n$-dimensional space of symmetric functions and \begin{equation*}
 \abs{f_n}^2_{\phi}=\!\! \int_{\Re^{2n}_+}\!\! f_n(s_1,\cdots,s_n)f_n(s'_1,\cdots,s'_n)\, \phi(s_1,s'_1)\cdots \phi(s_n,s'_n)\, ds_1\cdots ds_n\, ds'_1\cdots ds'_n .
\end{equation*}
$L^2_{\phi}(\Re^n_+)$ is the $n$-dimensional  space of symmetric functions. Given the base complete orthonormal base $\{\tilde{\xi}_n\}_{n\in\mathbb{N}}\subset L^2_\phi(\Re_+)$ then $L^2_{\phi}(\Re^n_+)$ is the completion of all function of the following form:
\begin{equation*}
f(s_1,\dots,s_n)= \sum_{1\leq k_1,\dots, k_n\leq k} a_{ k_1,\dots, k_n}\tilde{\xi}_{k_1}(s_1)\,\tilde{\xi}_{k_2}(s_2)\cdots\tilde{\xi}_{k_n}(s_n).
\end{equation*}
Associated to this functions we define the multiple integral as
\begin{equation*}
I_n(f)=\sum_{1\leq k_1,\dots, k_n\leq k} a_{ k_1,\dots, k_n} \int_\Re\tilde{\xi}_{k_1}(u)\, dB^H_u\diamond\,\int_\Re\tilde{\xi}_{k_2}(u)\, dB^H_u \diamond\cdots\diamond\int_\Re\tilde{\xi}_{k_n}(u)\, dB^H_u,
\end{equation*}
It is not difficult to prove \citep[\textbf{Lemma 6.6}]{paper:duncan} that given $f\in L^2_{\phi}(\Re^n_+)$ and $f\in L^2_{\phi}(\Re^m_+)$ it is
\begin{equation}
\E{I_n(f)I_m(g)}=\left\{\begin{aligned}
(f,g)_\phi &\quad\text{if }n=m\\
0\quad\quad &\quad \text{if }n\neq m\end{aligned}\right..
\label{eq:orthonorm-int}
\end{equation}
Moreover, for the iterated integral
\begin{multline*}
\int_{0\leq s_1\leq s_2<\cdots<s_n\leq t} f_n(s_1,\cdots,s_n)\> dB^H_{s_1}\cdots dB^H_{s_n}=\\
\int_0^t \left(\int_{0\leq s_1\leq s_2<\cdots<s_{n-1}\leq s_n}f_n(s_1,\cdots,s_{n-1},s_n)\> dB^H_{s_1}\cdots dB^H_{s_{n-1}}\right)dB^H_{s_n}
\end{multline*}
is $n!$ times $I_n(f)$.
\bibliography{bib-dar}
\end{document}